\documentclass[showkeys,showpacs,superscriptaddress]{revtex4}
\usepackage{amsmath}
\usepackage{amssymb}
\usepackage{graphicx}

\begin{document}

\title{
Proca tubes with the flux of the longitudinal chromoelectric field \\
and the energy flux/momentum density
}

\author{
Vladimir Dzhunushaliev
}
\email{v.dzhunushaliev@gmail.com}
\affiliation{
Department of Theoretical and Nuclear Physics,  Al-Farabi Kazakh National University, Almaty 050040, Kazakhstan
}
\affiliation{
Institute of Experimental and Theoretical Physics,  Al-Farabi Kazakh National University, Almaty 050040, Kazakhstan
}
\affiliation{
Academician J.~Jeenbaev Institute of Physics of the NAS of the Kyrgyz Republic, 265 a, Chui Street, Bishkek 720071, Kyrgyzstan
}

\author{Vladimir Folomeev}
\email{vfolomeev@mail.ru}
\affiliation{
Institute of Experimental and Theoretical Physics,  Al-Farabi Kazakh National University, Almaty 050040, Kazakhstan
}
\affiliation{
Academician J.~Jeenbaev Institute of Physics of the NAS of the Kyrgyz Republic, 265 a, Chui Street, Bishkek 720071, Kyrgyzstan
}
\affiliation{
International Laboratory for Theoretical Cosmology, Tomsk State University of Control Systems and Radioelectronics (TUSUR),
Tomsk 634050, Russia
}
\begin{abstract}
We consider non-Abelian SU(3) Proca theory with a Higgs scalar field included.
Cylindrically symmetric solutions describing classical tubes either with the flux of a longitudinal electric field or with the energy flux (and hence with nonzero momentum density) are obtained. It is shown that, in quantum Proca theory, there can exist tubes both with the flux of the longitudinal electric field and with the energy flux/momentum density simultaneously. An imaginary particle -- Proca proton~-- in which `quarks' are connected by tubes with nonzero momentum density is considered. It is shown that this results in the appearance of the angular momentum related to the presence of the non-Abelian electric and magnetic fields in the tube, and this angular momentum is a part of the Proca proton spin.
\end{abstract}



\date{\today}

\maketitle

\section{Introduction}
In recent years there is a growing interest in modelling systems containing various massive vector fields.
Such fields are described within gauge Proca theories (both Abelian and non-Abelian ones) where the gauge invariance
is violated explicitly by introducing a mass term.
Being the generalization of Maxwell's theory, Proca theory
permits one to take into account various effects related to the possible presence of the mass of vector particles.
At the present time, Proca fields are used in different aspects: in constructing models of black holes~\cite{Heisenberg:2017xda},
hypothetical Proca stars~\cite{Brito:2015pxa,Herdeiro:2017fhv,Dzhunushaliev:2019kiy,Herdeiro:2019mbz},
in studying the generalized Proca theories~\cite{Heisenberg:2014rta,Allys:2015sht,Heisenberg:2016eld} and solitons~\cite{Babichev:2017rti},
in describing the massive spin-1 $Z^0$ and $W^\pm$ bosons in the standard model~\cite{Lawrie2002}, in considering various effects related to the possible presence of the rest mass of a photon~\cite{Tu:2005ge},
and within dark matter physics~\cite{Pospelov:2008jd}.

When considering configurations containing Proca fields, the main efforts have been focused on studying spherically symmetric solutions.
However, a consideration of cylindrically  symmetric systems seems to be of interest as well.
In particular, cylindrically symmetric localized regular solutions (the so-called `Proca Q tubes') have been found in Ref.~\cite{Brihaye:2016pld}.

The next natural step in this direction is to generalize systems containing Proca fields by including extra fields in them.
For example, for a spherically symmetric case, in Proca theories involving some extra fields, there can occur topologically trivial monopoles with
an exponentially decaying radial magnetic field~\cite{Dzhunushaliev:2019ham}. In the present paper we show that, in non-Abelian
Proca theories containing a Higgs scalar field, there exist cylindrically symmetric solutions describing tubes filled with color electric and magnetic fields.
The distinctive feature of these tubes is that they can contain either the flux of the longitudinal electric field sourced by `quarks' located at
 $\pm \infty$ or the energy flux associated with the presence of nonzero Poynting vector (and hence a nonzero momentum density).
[Henceforth, by `quarks' (in quotation marks) we mean the sources of non-Abelian Proca fields, analogous to what is done in quantum chromodynamics
for Yang-Mills fields and real quarks.]

The paper is organized as follows. In Sec.~\ref{Proca_Dirac_scalar}, we write down the general field equations for the non-Abelian-Proca-Higgs theory.
In Sec.~\ref{QMplusQuarks}, we obtain cylindrically symmetric solutions to the equations of Sec.~\ref{Proca_Dirac_scalar} describing tubes which
contain either the flux of the electric field or the energy flux (and hence the momentum density). In Sec.~\ref{momentum_electric_tube}, we present some qualitative arguments in favor of the fact that, in quantum Proca theory, there can exist tubes containing simultaneously both
the flux of the longitudinal electric field and the energy flux/momemtum density.
In Sec.~\ref{Proca_angular_momentum}, we show that tubes with nonzero momentum density connecting three `quarks' in a Proca proton do create the angular momentum that is a part of the Proca proton spin.
Finally, in Sec.~\ref{concl}, we summarize the results obtained in the present paper.

\section{Non-Abelian-Proca-Higgs theory}
\label{Proca_Dirac_scalar}

The Lagrangian describing a system consisting of a non-Abelian SU(3) Proca field $A^a_\mu$ interacting with nonlinear scalar field $\phi$ can be taken in the form
(hereafter, we work in units such that $c=\hbar=1$)
\begin{equation}
	\mathcal L =  - \frac{1}{4} F^a_{\mu \nu} F^{a \mu \nu} -
	\frac{\left( \mu^2 \right)^{a b, \mu}_{\phantom{a b,}\nu}}{2}
	A^a_\mu A^{b \nu} +
	\frac{1}{2} \partial_\mu \phi \partial^\mu \phi +
	\frac{\lambda}{2} \phi^2 A^a_\mu A^{a \mu} -
	\frac{\Lambda}{4} \left( \phi^2 - M^2 \right)^2.
\label{2_10}
\end{equation}
Here
$
	F^a_{\mu \nu} = \partial_\mu A^a_\nu - \partial_\nu A^a_\mu +
	g f_{a b c} A^b_\mu A^c_\nu
$ is the field strength tensor for the Proca field, where $f_{a b c}$ are the SU(3) structure constants, $g$ is the coupling constant,
$a,b,c = 1,2, \dots, 8$ are color indices,
$\mu, \nu = 0, 1, 2, 3$ are spacetime indices. The Lagrangian \eqref{2_10} also contains the arbitrary constants $M, \lambda, \Lambda$ and the Proca field mass matrix
$
	\left( \mu^2 \right)^{a b, \mu}_{\phantom{a b,}\nu}
$.

Using \eqref{2_10}, the corresponding field equations can be written in the form
\begin{eqnarray}
	D_\nu F^{a \mu \nu} - \lambda \phi^2 A^{a \mu} &=&
	- \left( \mu^2 \right)^{a b, \mu}_{\phantom{a b,}\nu} A^{b \nu},
\label{2_20}\\
	\Box \phi &=& \lambda A^a_\mu A^{a \mu} \phi +
	\Lambda \phi \left( M^2 - \phi^2 \right) ,
\label{2_30}
\end{eqnarray}
and the energy density is
\begin{equation}
\begin{split}
	\varepsilon = &\frac{1}{2} \left( E^a_i \right)^2 +
	\frac{1}{2} \left( H^a_i \right)^2 -
	\left[
		\left( \mu^2 \right)^{a b, \alpha}_{\phantom{a b,} 0} A^a_\alpha A^b_0 -
		\frac{1}{2} \left( \mu^2 \right)^{a b, \alpha}_{\phantom{a b,} \beta} A^a_\alpha A^{b \beta}
	\right]
	+
	\frac{1}{2} \left( \partial_t \phi \right)^2 +
	\frac{1}{2} \left( \nabla \phi \right)^2
\\
	&
	+\lambda \phi^2 \left[
		\left( A^a_0 \right)^2 - \frac{1}{2} A^a_\alpha A^{a \alpha}
	\right] +
	\frac{\Lambda}{4} \left( \phi^2 - M^2 \right)^2 ,
\label{2_40}
\end{split}
\end{equation}
where $i=1,2,3$ and $E^a_i$ and $H^a_i$ are the components of the electric and magnetic field strengths, respectively.

\section{Proca flux tubes with the energy flux (momentum density) and the flux of the electric field}
\label{QMplusQuarks}

The focus of this section is on obtaining and solving equations for two different types of flux tubes 
containing either the flux of the electric field or the energy flux (and hence the momentum density).

\subsection{Proca tube with the flux of the longitudinal electric field}
\label{electric_flux}

To obtain a tube filled with a longitudinal color electric field, we choose the {\it Ans\"{a}tze}
\begin{equation}
	A^2_t = \frac{h(\rho)}{g} , \;
	A^5_z = \frac{v(\rho)}{g} , \;
	A^7_\varphi = \frac{\rho w(\rho)}{g} , \;
 	\phi = \phi(\rho),
\label{3_a_10}
\end{equation}
where $\rho, z,$ and $\varphi$ are cylindrical coordinates. 
These {\it Ans\"{a}tze} follow from the general cylindrically symmetric {\it Ansatz} of Ref.~\cite{Singleton:1995xc, Obukhov:1996ry}, where it is shown that it is consistent with Yang-Mills-Higgs theory.
For such a choice of the  SU(3) Proca field potentials, we have the following electric and magnetic field intensities:
\begin{eqnarray}
	E^2_\rho &=& - \frac{h^\prime}{g} , \quad
	E^5_\varphi = - \frac{\rho h w}{2 g} , \quad
	E^7_z = \frac{h v}{2 g} ,
\label{3_a_20}\nonumber \\
H^2_\rho &=& - \frac{v w}{2 g}, \quad
	H^5_\varphi = - \frac{\rho v^\prime}{g}, \quad
	H^7_z = \frac{1}{g} \left(
		w^\prime + \frac{w}{\rho}
	\right) .
\label{3_a_30}
\nonumber
\end{eqnarray}
In this case the energy flux is absent, since the Poynting vector is zero,
\begin{equation}
	S^i = \frac{\epsilon^{i j k}}{\sqrt{\gamma}} E^a_j H^a_k = 0.
\label{3_a_40}
\end{equation}
Here, $\epsilon^{i j k}$ is the completely antisymmetric Levi-Civita symbol and $\gamma$ is the determinant of the space metric.
For such a tube, the energy density \eqref{2_40} yields
\begin{equation}
\begin{split}
	g^2 \varepsilon = & \frac{\left( h^\prime \right)^2}{2} +
	\frac{\left( v^\prime \right)^2}{2} +
	\frac{1}{2} \left(
		w^\prime + \frac{w}{\rho}
	\right)^2 +
	g^2 \frac{\left( \phi^\prime \right)^2}{2} +
	\frac{1}{8} \left( w^2 h^2 + w^2 v^2 + h^2 v^2 \right) -
	\frac{\mu_1^2}{2} h^2 - \frac{\mu_2^2}{2} v^2 -
	\frac{\mu_3^2}{2}  w^2
\\
	+&
	\frac{\lambda}{2} \phi^2 \left(
		h^2 + v^2 + w^2
	\right) +
			\frac{g^2 \Lambda}{4} \left( \phi^2 - M^2 \right)^2
\end{split}
\label{3_a_50}
\end{equation}
with the following components of the Proca field mass matrix: $\mu_{1}^2=\left( \mu^2 \right)^{2 2, t}_{\phantom{a b,}t}$,
 $\mu_{2}^2=\left( \mu^2 \right)^{5 5, z}_{\phantom{a b,}z}$, and $\mu_{3}^2=\left( \mu^2 \right)^{7 7, \varphi}_{\phantom{a b,}\varphi}$.

Substituting the potentials \eqref{3_a_10} in Eqs.~\eqref{2_20} and \eqref{2_30} and introducing the dimensionless variables
 $\tilde \phi = \phi \sqrt{\lambda} / \phi(0)$,
$\tilde h = h / \phi(0)$,
$\tilde v = v / \phi(0)$,
$\tilde w = w / \phi(0)$,
$\tilde M = M \sqrt{\lambda} / \phi(0)$,
$\tilde \lambda = \lambda / g^2$,
$\tilde \Lambda = \Lambda / \lambda$,
$\tilde \mu_{1,2,3} = \mu_{1,2,3} / \phi(0)$, and
$x = \rho \phi(0)$ [here $\phi(0)$ is the central value of the scalar field], we get the following set of equations:
\begin{eqnarray}
  \tilde h'' + \frac{\tilde h'}{x} &=& h
  \left(
  	\frac{{\tilde v}^2}{4} + \frac{\tilde w^2}{4} + {\tilde \phi}^2 -
  	\tilde \mu_1^2
  \right) ,
\label{3_a_93}\\
  \tilde v'' + \frac{\tilde v'}{x} &=& v
  \left(
	  - \frac{\tilde h^2}{4} + \frac{\tilde w^2}{4} + {\tilde \phi}^2 -
	  \tilde \mu_2^2
  \right) ,
\label{3_a_96}\\
  \tilde w'' + \frac{\tilde w'}{x} - \frac{\tilde w}{x^2} &=&
  \tilde w \left(
  	- \frac{\tilde h^2}{4} + \frac{\tilde v^2}{4} + {\tilde \phi}^2 -
  	\tilde \mu_3^2
  \right),
\label{3_a_97}\\
  \tilde \phi'' + \frac{\tilde \phi'}{x} &=&
  \tilde \phi
  \left[
  	\tilde \lambda \left( - \tilde h^2 + \tilde v^2 + \tilde w^2 \right) +
  	\tilde \Lambda \left( {\tilde \phi}^2 - {\tilde M}^2 \right)
  \right].
\label{3_a_99}
\end{eqnarray}
Here,  the prime denotes differentiation with respect to the dimensionless radius $x$.
We seek a solution to Eqs.~\eqref{3_a_93}-\eqref{3_a_99} in the vicinity of the origin of coordinates in the form
\begin{eqnarray}
	\tilde h(x) &=& \tilde h_0 + \tilde h_2 \frac{x^2}{2} + \dots \quad \text{with} \quad
\tilde h_2 = \frac{\tilde h_0}{2} \left(
		\frac{\tilde v_0^2}{4} + \tilde \phi_0^2 - \tilde \mu^2_1
	\right) ,
\label{3_a_100}\nonumber\\	
	\tilde v(x) &=& \tilde v_0 + \tilde v_2 \frac{x^2}{2} + \dots \quad \text{with} \quad
\tilde v_2 = \frac{\tilde v_0}{2} \left(
		- \frac{\tilde h_0^2}{4} + \tilde \phi_0^2 - \tilde \mu^2_2
	\right) ,
\label{3_a_110}\nonumber\\
	\tilde w(x) &=& \tilde w_1 x + \tilde w_3 \frac{x^3}{3 !} + \dots \quad \text{with} \quad
\tilde w_3 = \frac{3}{4}\tilde w_1 \left(
		- \frac{\tilde h_0^2}{4} + \frac{\tilde v_0^2}{4} +
		\tilde \phi_0^2 - \tilde \mu^2_3
	\right) ,
\label{3_a_115}\nonumber\\
	\tilde \phi(x) &=& \tilde \phi_0 + \tilde \phi_2 \frac{x^2}{2} + \dots \quad \text{with} \quad
\tilde \phi_2 = \frac{\tilde \phi_0}{2} \left[
		\tilde \lambda \left( - \tilde h_0^2 + \tilde v_0^2 \right) +
		\tilde \Lambda \left(
			{\tilde \phi}_0^2 - {\tilde M}^2
	\right)
	\right],
\label{3_a_120}
\nonumber
\end{eqnarray}
where the expansion coefficients $\tilde h_0, \tilde v_0, \tilde \phi_0$, and $\tilde w_1$ are arbitrary.

\begin{figure}[t]
\begin{minipage}[t]{.49\linewidth}
	\begin{center}
		\includegraphics[width=1\linewidth]{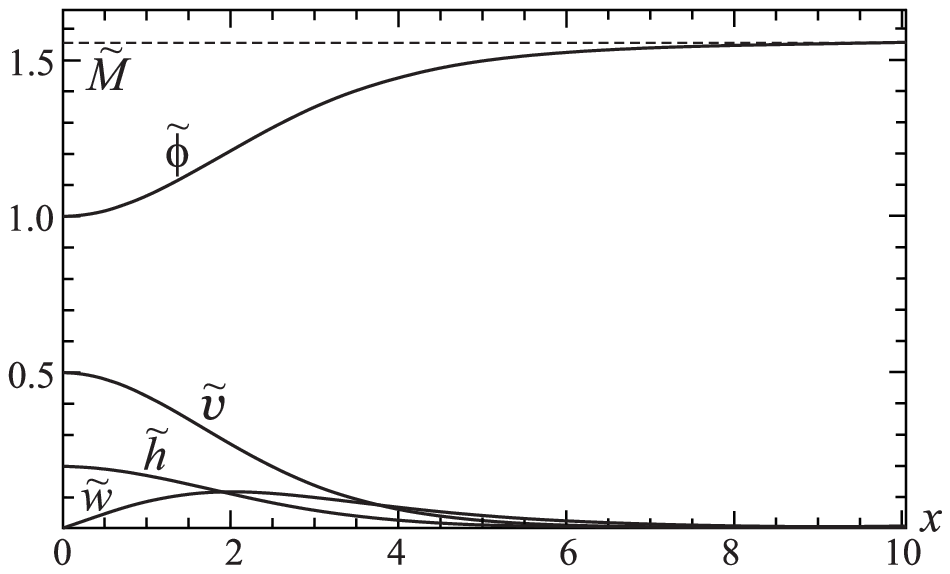}
	\end{center}
\vspace{-0.5cm}		
\caption{The graphs of the Proca field potentials
$\tilde h\equiv g A^2_t / \phi(0)  $, $\tilde v\equiv g A^5_z / \phi(0) $,
$\tilde w\equiv g A^7_\varphi / x $ and of the scalar field $\tilde \phi$ for the following values of the system parameters:
$
	\tilde \lambda = 2, \tilde \Lambda = 0.1, \tilde h_0 = 0.2, \tilde v_0 = 0.5,
	\tilde \phi_0 = 1, \tilde w_1 = 0.1,
	\tilde M = 1.5520735, \tilde \mu_1 =1.29428281 , \tilde \mu_2 = 1.28105328, \tilde \mu_3 = 1.469883.
$
}
\label{h_v_w_phi}
\end{minipage}\hfill
\begin{minipage}[t]{.49\linewidth}
	\begin{center}
		\includegraphics[width=1\linewidth]{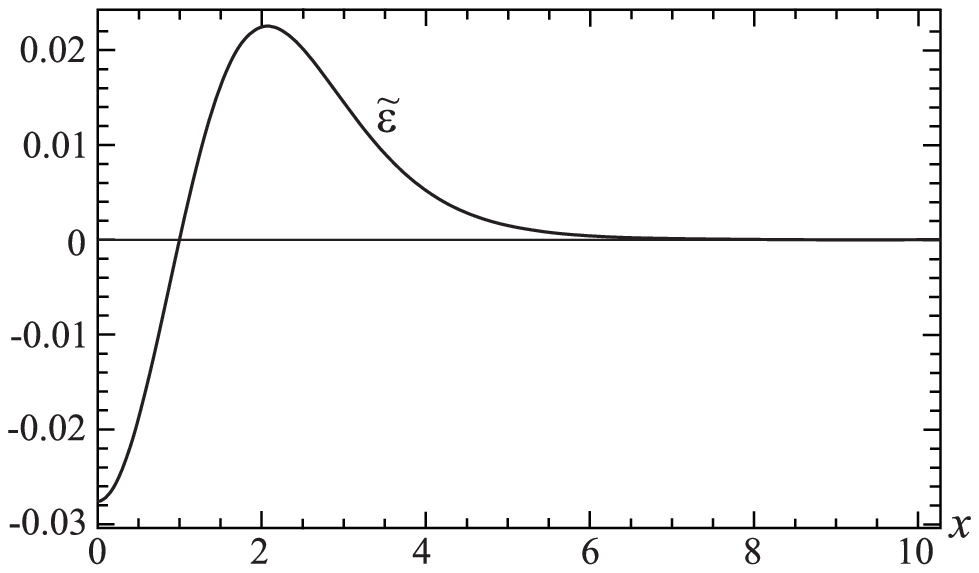}
	\end{center}
\vspace{-0.5cm}
\caption{The profile of the dimensionless flux tube energy density $\tilde \varepsilon\equiv g^2 \varepsilon/\phi^4(0)$ from
 Eq.~\eqref{3_a_50} rewritten in terms of the dimensionless variables given before Eq.~\eqref{3_a_93}. 	
}
\label{energy_density}
\end{minipage} \\
\begin{minipage}[t]{.49\linewidth}
	\begin{center}
		\includegraphics[width=1\linewidth]{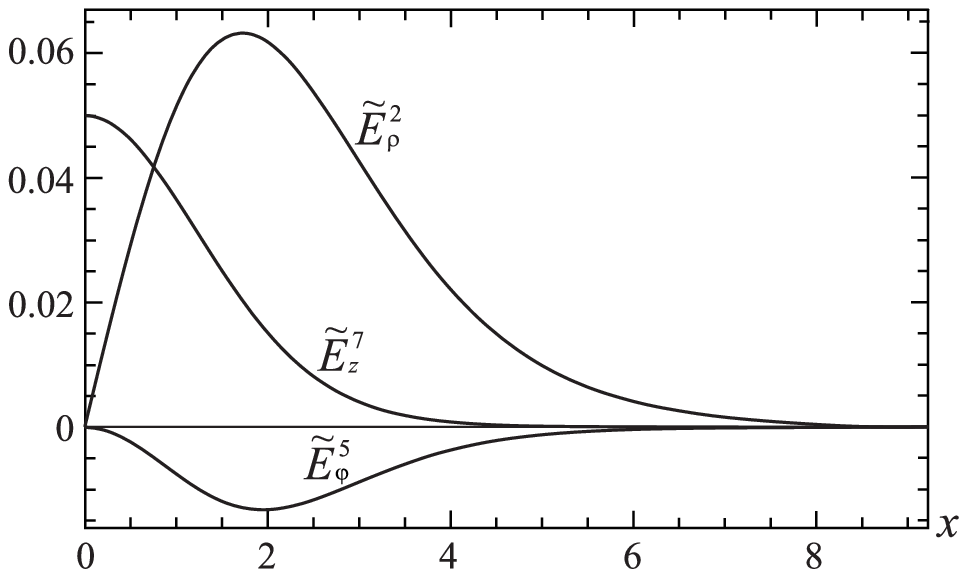}
	\end{center}
\vspace{-0.5cm}
\caption{The profiles of the color electric fields	
$\tilde E^2_\rho\equiv g E^2_\rho / \phi^2(0), \tilde E^5_\varphi \equiv g E^5_\varphi / \phi(0)$, and $\tilde E^7_z\equiv  g E^7_z / \phi^2(0)$.
}
\label{electric_fields}
\end{minipage} \hfill
\begin{minipage}[t]{.49\linewidth}
	\begin{center}
		\includegraphics[width=1\linewidth]{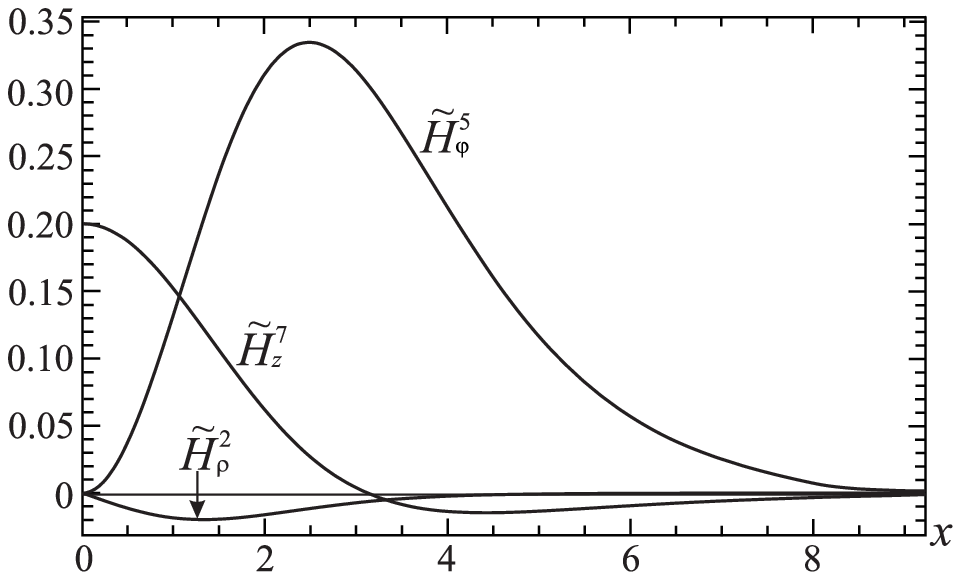}
	\end{center}
\vspace{-0.5cm}
\caption{The profiles of the color magnetic fields	
$\tilde H^2_\rho\equiv g H^2_\rho / \phi^2(0), \tilde H^5_\varphi\equiv g H^5_\varphi / \phi(0)$, and $\tilde H^7_z\equiv  g H^7_z / \phi^2(0)$.
}
\label{magnetic_fields}
\end{minipage} \hfill
\end{figure}

The derivation of solutions to the set of equations~\eqref{3_a_93}-\eqref{3_a_99} is an eigenvalue problem for the parameters $\tilde \mu_{1}, \tilde \mu_{2}, \tilde \mu_{3}$, and $\tilde M$.
The numerical solution describing the behavior of the Proca field potentials and of the corresponding electric and magnetic fields is given in Figs.~\ref{h_v_w_phi}-\ref{magnetic_fields}.

The asymptotic behavior of the functions $\tilde h, \tilde v, \tilde w$, and $\tilde \phi$, which follows from Eqs.~\eqref{3_a_93}-\eqref{3_a_99}, is
\begin{equation}
	\tilde h(x) \approx \tilde h_{\infty}
	\frac{e^{- x \sqrt{\tilde M^2 - \tilde \mu^2_1}}}{\sqrt x},\,\,
	\tilde v(\rho) \approx \tilde v_{\infty}
	\frac{e^{- x \sqrt{\tilde M^2 - \tilde \mu^2_2}}}{\sqrt x},\,\,
	\tilde w(\rho) \approx \tilde w_{\infty}
	\frac{e^{- x \sqrt{\tilde M^2 - \tilde \mu^2_3}}}{\sqrt x},\,\,
	\tilde \phi \approx \tilde M - \tilde \phi_\infty
	\frac{e^{- x \sqrt{2 \tilde \Lambda \tilde M^2}}}{\sqrt x} ,
\label{3_a_160}
\nonumber
\end{equation}
where $\tilde h_{\infty}, \tilde v_{\infty}, \tilde w_{\infty}$, and $\tilde \phi_\infty$ are integration constants.

This  flux tube solution has the following characteristics:
\begin{itemize}
\item All physical quantities, such as the Proca field potentials, the electric and magnetic field intensities, and the energy density, are finite.
\item The tube contains the finite flux of the longitudinal color electric field $E^7_z$,
$$
	\Phi_z = 2 \pi \int \limits_0^\infty \rho E^7_z d \rho < \infty .
$$
\item The tube possesses the finite linear energy density, 
$$
	\mathcal{E} = 2 \pi \int \limits_0^\infty \rho\, \varepsilon(\rho) d \rho
	< \infty .
$$
\end{itemize}

Thus in this section we have demonstrated that there are cylindrically symmetric solutions within the non-Abelian-Proca-Higgs theory.
These solutions describe a tube with the flux of the longitudinal color electric field, and this field is sourced by `quarks'/`antiquarks'
located at $z=\pm \infty$.

\subsection{Proca tube with the energy flux/momentum density}
\label{momentum_flux}

Here, we choose the {\it Ans\"{a}tze}
\begin{equation}
	A^5_t = \frac{f(\rho)}{g} , \;
	A^5_z = \frac{v(\rho)}{g} , \;
	A^7_\varphi = \frac{\rho w(\rho)}{g} , \;
 	\phi = \phi(\rho),
\label{3_b_10}
\end{equation}
 which give the following components of the electric and magnetic field intensities:
\begin{eqnarray}
	E^2_\varphi &=& \frac{\rho f w}{2 g}, \quad
	E^5_\rho = - \frac{f^\prime}{g} , \quad
\label{3_b_20}\\
	H^2_\rho &=& - \frac{v w}{2 g}, \quad
	H^5_\varphi = - \frac{\rho v^\prime}{g}, \quad
	H^7_z = \frac{1}{g} \left(
		w^\prime + \frac{w}{\rho}
	\right) .
\label{3_b_30}
\end{eqnarray}
In this case the Poynting vector~\eqref{3_a_40} is already nonzero
[see Eq.~\eqref{3_b_50} below].

Substituting the potentials \eqref{3_b_10} in Eqs.~\eqref{2_20} and \eqref{2_30} and using the dimensionless variables
 given before Eq.~\eqref{3_a_93}, we derive the following equations:
\begin{eqnarray}
  \tilde f'' + \frac{\tilde f'}{x} &=& \tilde f
  \left(
  	\frac{\tilde w^2}{4} + {\tilde \phi}^2 - \tilde \mu_1^2
  \right) ,
\label{3_b_100}\\
  \tilde v'' + \frac{\tilde v'}{x} &=& \tilde v
  \left(
	  \frac{\tilde w^2}{4} + {\tilde \phi}^2 - \tilde \mu_2^2
  \right) ,
\label{3_b_110}\\
  \tilde w'' + \frac{\tilde w'}{x} - \frac{\tilde w}{x^2} &=&
  \tilde w \left(
  	- \frac{\tilde f^2}{4} + \frac{\tilde v^2}{4} + {\tilde \phi}^2 -
  	\tilde \mu_3^2
  \right),
\label{3_b_120}\\
  \tilde\phi'' + \frac{\tilde \phi'}{x} &=&
  \tilde \phi
  \left[
  	\tilde \lambda \left( - \tilde f^2 + \tilde v^2 + \tilde w^2 \right) +
  	\tilde \Lambda \left( {\tilde \phi}^2 - {\tilde M}^2 \right)
  \right],
\label{3_b_130}
\end{eqnarray}
where $\tilde f = f / \phi(0)$.

This set of equations has a cylindrically symmetric solution describing a tube with nonzero momentum density and energy flux (the Poynting vector).
To demonstrate this, let us consider the simplest particular case where
$\tilde f = \tilde v$ and $\tilde \mu_1 = \tilde \mu_2 = \tilde \mu$. In this case the set of equations
\eqref{3_b_100}-\eqref{3_b_130} is split as follows. Eq.~\eqref{3_b_100} takes the form of the Schr\"{o}dinger equation,
\begin{equation}
  - \tilde f'' - \frac{\tilde f'}{x} + \tilde f U_{\tilde f, \text{eff}}	=
  \tilde \mu^2 \tilde f,
\label{3_b_140}
\end{equation}
where the effective potential for the ``wave function'' $\tilde f$ is
\begin{equation}
	U_{\tilde f, \text{eff}} = \frac{\tilde w^2}{4} + {\tilde \phi}^2 .
\label{3_b_150}
\end{equation}
In order to ensure a regular solution of this equation, it is necessary that the effective potential would possess a well.
In this case Eq.~\eqref{3_b_140} must be solved as an eigenvalue problem for the parameter
$\tilde \mu^2$ with the eigenfunction $\tilde f$.

The remaining equations \eqref{3_b_120} and \eqref{3_b_130} for the functions
$\tilde w$ and $\tilde \phi$ are then
\begin{eqnarray}
  - \tilde w'' - \frac{\tilde w'}{x} + \tilde w U_{\tilde w, \text{eff}} &=& \tilde \mu_3^2 \tilde w ,
\label{3_b_160}\\
  \tilde \phi'' + \frac{\tilde \phi'}{x} &=&
  \tilde \phi
  \left[
  	\tilde \lambda \tilde w^2 +
  	\tilde \Lambda \left( {\tilde \phi}^2 - {\tilde M}^2 \right)
  \right],
\label{3_b_170}
\end{eqnarray}
and they do not already contain the function $\tilde f$. The effective potential, which appears in Eq.~\eqref{3_b_160}, is
\begin{equation}
	U_{\tilde w, \text{eff}} = \frac{1}{x^2} + {\tilde \phi}^2.
\label{3_b_180}
\end{equation}

The equation~\eqref{3_b_160} has the form of the Schr\"{o}dinger equation with the ``wave function''
 $\tilde w$ and with the ``energy'' $\tilde \mu_3^2$. This means that it will have a regular solution only if the effective potential
 $U_{\tilde w, \text{eff}}$ possesses a well. Note also that in the limit
$x\to 0$ the effective potential $U_{\tilde w, \text{eff}} \to 1/x^2$, i.e., it is repulsive and hence the ``fall of a particle to the centre'' is certainly absent.

We will seek regular solutions possessing a finite linear energy density. This means that asymptotically (as $x \rightarrow \infty$) the functions behave as
$
	\tilde f(x), \tilde v(x), \tilde w(x) \rightarrow 0
$.
Then, taking into account the positiveness of the effective potentials \eqref{3_b_150} and \eqref{3_b_180}, one can conclude that the function
$\tilde \phi$ must go to a constant, and Eq.~\eqref{3_b_170} implies that this constant is $\tilde M$, i.e.,
$
	\tilde \phi \rightarrow \tilde M	\text{ as } x \rightarrow \infty
$.

We seek solutions of Eqs.~\eqref{3_b_140}, \eqref{3_b_160}, and \eqref{3_b_170} in the vicinity of the origin of coordinates in the form
\begin{eqnarray}
	\tilde f(x) &=& \tilde f_0 + \tilde f_2 \frac{x^2}{2} + \dots \quad \text{with} \quad \tilde f_2 = \frac{\tilde f_0}{2} \left(
		 \tilde \phi_0^2 - \tilde \mu^2
	\right) ,
\label{3_b_205}\nonumber\\	
	\tilde w(x) &=& \tilde w_1 x + \tilde w_3 \frac{x^3}{3 !} + \dots \quad \text{with} \quad \tilde w_3 = \frac{2}{3}\tilde w_1 \left(\tilde \phi_0^2 -\tilde \mu_3^2 \right),
\label{3_b_210}\nonumber\\
	\tilde \phi(x) &=& \tilde \phi_0 + \tilde \phi_2 \frac{x^2}{2} + \dots \quad \text{with} \quad
\tilde \phi_2 = \frac{\tilde \phi_0}{2} \tilde \Lambda \left(
		{\tilde \phi}_0^2 - {\tilde M}^2 \right),
\label{3_b_220}
\nonumber
\end{eqnarray}
where the expansion coefficients $\tilde f_0, \tilde \phi_0$, and $\tilde w_1$ are arbitrary.

In turn, the asymptotic behavior of the functions is
$$
	\tilde f(x) = \tilde v(x) \approx f_{\infty}
	\frac{e^{- x \sqrt{\tilde M^2 - \tilde \mu^2}}}{\sqrt x},\quad
	\tilde w(x) \approx \tilde w_{\infty}
	\frac{e^{- x \sqrt{\tilde M^2 - \tilde \mu^2_3}}}{\sqrt x},\quad
	\tilde \phi \approx \tilde M - \tilde \phi_\infty
	\frac{e^{- x \sqrt{2 \tilde \Lambda \tilde M^2}}}{\sqrt x} ,
$$
where $f_{\infty}, \tilde w_{\infty}$, and $\tilde \phi_\infty$ are integration constants.

\begin{figure}[t]
\begin{minipage}[t]{.49\linewidth}
	\begin{center}
		\includegraphics[width=1\linewidth]{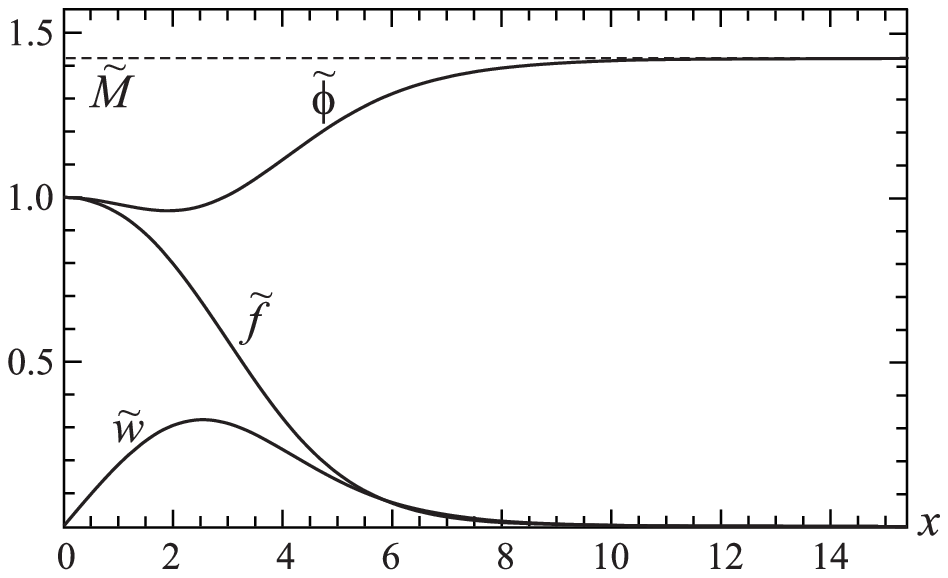}
	\end{center}
\vspace{-0.5cm}		
\caption{The graphs of the functions $\tilde w, \tilde \phi$, and $\tilde f$
for the following values of the system parameters:
$
	\tilde \lambda = 2, \tilde \Lambda = 0.1,
	\tilde \phi_0 = 1, \tilde w_1 = 0.2, \tilde f_0 = 1,
	\tilde M = 1.42141, \tilde \mu = 1.0876563, \tilde \mu_3 = 1.206829.
$
}
\label{w_phi_f}
\end{minipage}\hfill
\begin{minipage}[t]{.49\linewidth}
	\begin{center}
		\includegraphics[width=1\linewidth]{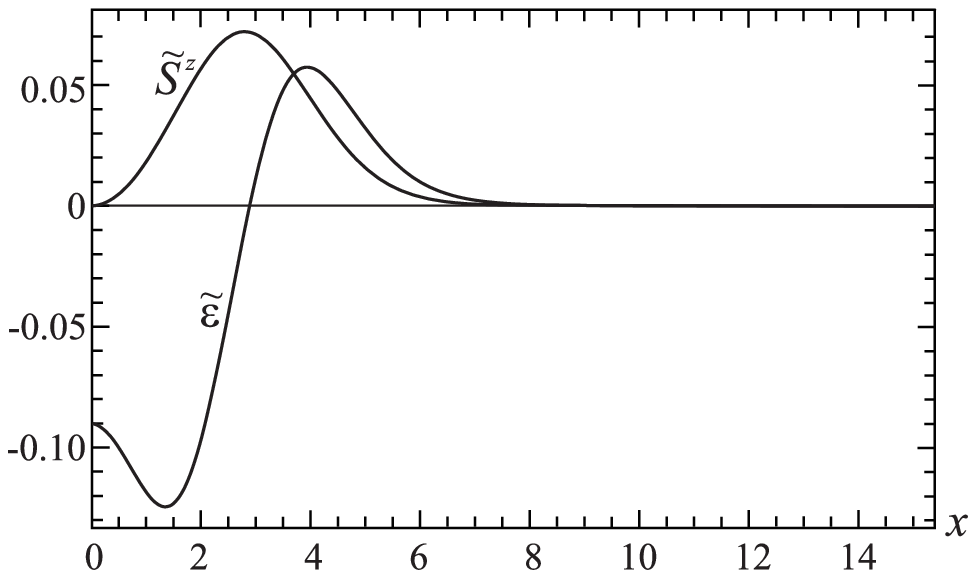}
	\end{center}
\vspace{-0.5cm}
\caption{The profiles of the $z$-component of the Poynting vector $\tilde S^z $ from Eq.~\eqref{3_b_50}
and of the energy density $\tilde \varepsilon$ from Eq.~\eqref{3_b_260}.
}
\label{Poynting}
\end{minipage} \\
\begin{minipage}[t]{.49\linewidth}
	\begin{center}
		\includegraphics[width=1\linewidth]{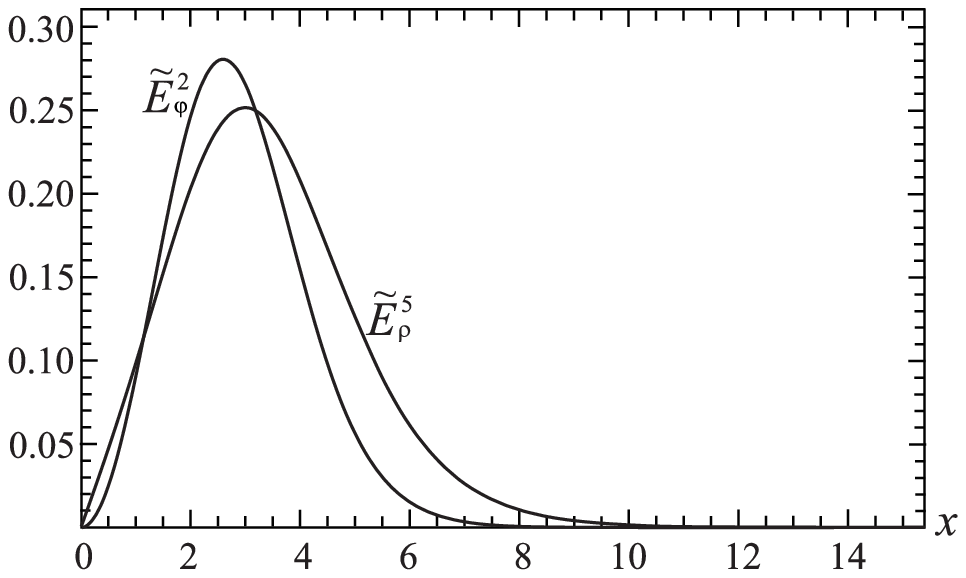}
	\end{center}
\vspace{-0.5cm}
\caption{The profiles of the color electric fields
$\tilde E^2_\varphi\equiv g E^2_\varphi / \phi(0)$ and $\tilde E^5_\rho\equiv g E^5_\rho / \phi^2(0)$. 	
}
\label{el_fields}
\end{minipage}\hfill
\begin{minipage}[t]{.49\linewidth}
	\begin{center}
		\includegraphics[width=1\linewidth]{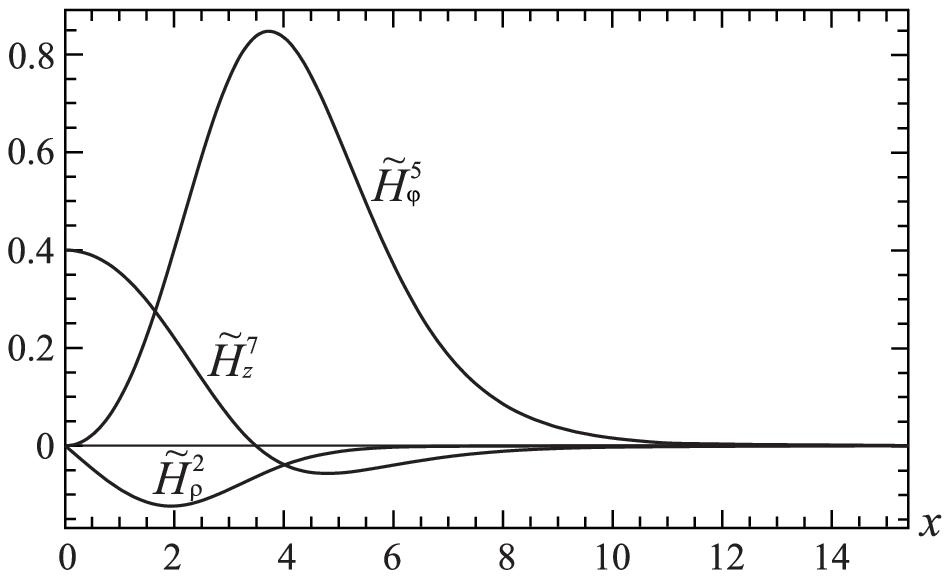}
	\end{center}
\vspace{-0.5cm}
\caption{The profiles of 	the color magnetic fields
$\tilde H^2_\rho\equiv g H^2_\rho / \phi^2(0), \tilde H^5_\varphi\equiv g H^5_\varphi / \phi(0)$, and $\tilde H^7_z\equiv g H^7_z / \phi^2(0)$.
}
\label{magn_fields}
\end{minipage}
\end{figure}

The results of numerical calculations are shown in Figs.~\ref{w_phi_f}-\ref{magn_fields}, including the graph for the dimensionless energy density
 \begin{equation}
\begin{split}
	\tilde \varepsilon\equiv \frac{g^2}{\phi^4(0)}  \varepsilon = &
	\frac{\left( \tilde f^\prime \right)^2}{2} +
	\frac{\left( \tilde v^\prime \right)^2}{2} +
	\frac{1}{2} \left(
		\tilde w^\prime + \frac{\tilde w}{x}
	\right)^2 +
	\frac{1}{\tilde \lambda} \frac{\left( \tilde{\phi}^\prime \right)^2}{2} +
	\frac{\tilde w^2}{8} \left( \tilde f^2 + \tilde v^2 \right) -
	\frac{\tilde \mu_1^2}{2} \tilde f^2 -
	\frac{\tilde \mu_2^2}{2} \tilde v^2 -
	\frac{\tilde \mu_3^2}{2} \tilde w^2
\\
	&
	+\frac{{\tilde \phi}^2}{2} \left(
		 \tilde f^2 + \tilde v^2 + \tilde w^2
	\right) +
			\frac{\tilde \Lambda}{4 \tilde \lambda}
			\left( \tilde \phi^2 - \tilde M^2 \right)^2
\end{split}
\label{3_b_260}
\end{equation}
given in Fig.~\ref{Poynting}.

Substitution of the components of electric and magnetic fields \eqref{3_b_20} and \eqref{3_b_30} in Eq.~\eqref{3_a_40}
yields the following expression for the Poynting vector:
\begin{equation}
	S^z = \frac{1}{g^2} \left(
		\frac{d f}{d \rho} \frac{d v}{d \rho} + \frac{1}{4}f v w^2
	\right)  =
	\frac{ \phi^4(0)}{g^2} \left(
		\tilde f^\prime \tilde v^\prime +
		\frac{1}{4}\tilde f \tilde v \tilde w^2
	\right) =
	\frac{ \phi^4(0)}{g^2} \tilde S^z .
\label{3_b_50}
\end{equation}
Note the presence of the gradient term $\tilde f^\prime \tilde v^\prime$, which is the same as that in Maxwell's electrodynamics,
and of the nonlinear term
$\tilde f \tilde v \tilde w^2/4$, which appears because the Proca field is non-Abelian.

The flux tube solution obtained in this section has the following characteristics:
\begin{itemize}
\item All physical quantities (the Proca electric and magnetic field potentials and intensities, the energy density, the momentum density, and the energy flux) are finite.
\item The tube contains a finite linear energy flux directed along the tube and computed as
$$
	\Pi^z = 2 \pi \int \limits_0^\infty \rho S^z(\rho) d \rho 	< \infty .
$$
\item Since the linear momentum density  $\vec{\mathcal P}$ is proportional to
the linear energy flux, $\vec{\mathcal P} = \vec \Pi $,
the tube also contains the $z$-component of the linear momentum density
$
	{\mathcal P}^z = \Pi^z
$.
\item In the tube, there is the finite flux of the longitudinal color magnetic field $H^7_z$,
$$
	\Phi_z = 2 \pi \int \limits_0^\infty \rho H^7_z d \rho < \infty .
$$
\item The tube possesses the finite linear energy density,
$$
	\mathcal{E} = 2 \pi \int \limits_0^\infty \rho \varepsilon(\rho) d \rho
	< \infty .
$$
\end{itemize}

Thus in this section we have shown that, in the non-Abelian-Proca-Higgs theory, there exist cylindrically symmetric solutions describing
a tube possessing the energy flux (and hence the momentum density) between `quark' and 'antiquark' located at $z=\pm \infty$.

\section{Proca flux tubes with a classical (quantum) flux of the electric field and quantum (classical) energy flux/momentum density}
\label{momentum_electric_tube}

In Sec.~\ref{QMplusQuarks}, we have shown that, within the non-Abelian-Proca-Higgs theory, there are two different types of tubes.
The tubes of the first type contain the flux of the longitudinal color electric field between `quarks' located at $\pm \infty$,
but there is no energy flux (and hence the momentum density). The tubes of the second type contain the energy flux/momentum density, but a flux of the longitudinal color electric field is absent.

In this connection, the natural question arises as to whether a tube possessing simultaneously both the flux of the longitudinal electric field
and the energy flux/momentum density does exist within classical non-Abelian-Proca-Higgs theory?
It seems to us that solutions describing such tubes are absent in classical theory,
but apparently they might exist in quantum non-Abelian-Proca-Higgs theory. In this section we give arguments in favor of such a point of view.

In Sec.~\ref{electric_flux}, we have shown that, in having the Proca field potentials  $A^2_t, A^5_z$, and $A^7_\varphi$ given in the form~\eqref{3_a_10},
one can obtain the longitudinal electric field $E^7_z$ (and hence the flux of this field between `quarks' which is needed to ensure confinement)
that occurs due to the nonlinearity of the field intensity tensor.
Similarly, in Sec.~\ref{momentum_flux}, we have shown that,  in having the Proca field potentials $A^5_t, A^5_z,$ and $A^7_\varphi$ given in the form~\eqref{3_b_10},
the energy flux/momentum density is present due to the appearance of the electric and magnetic fields $E^2_\varphi, H^2_\rho$ and $E^5_\rho, H^5_\varphi$
which are perpendicular to each other. A comparison of these two {\it Ans\"{a}tze} leads to the conclusion that for the simultaneous existence  of the flux
of the longitudinal electric field and of the energy flux/momentum density one needs the potentials $A^2_t, A^5_t, A^5_z,$ and $A^7_\varphi$.

We suppose that this can be done in two different ways, but only in quantum Proca theory.

\subsection{Tube with a classical flux of the longitudinal electric field and quantum energy flux/momentum density}
\label{ft_electric_field_momentum}
Here, the potentials are
$$
	\left\langle \hat{A}^2_t \right\rangle \neq 0 , \quad
	\left\langle \hat{A}^5_z \right\rangle \neq 0 , \quad
	\left\langle \hat{A}^7_\varphi \right\rangle \neq 0 , \quad
	\left\langle \hat{A}^5_t \right\rangle = 0,
$$
where $\left\langle \ldots \right\rangle$ denotes the quantum average. In this case, to describe the behavior of the quantum average of
$\hat{A}^2_t, \hat{A}^5_z$, and $\hat{A}^7_\varphi$, one can approximately use Eqs.~\eqref{3_a_93}-\eqref{3_a_99}.

Then, according to the definition of the Poynting vector \eqref{3_a_40}, there is the following quantum average:
\begin{equation}
	 \left\langle \widehat{S}^z \right\rangle  =
	\frac{1}{\sqrt \gamma} \left(
		\left\langle \widehat{E}^2_\varphi \widehat{H}^2_\rho \right\rangle +
		\left\langle \widehat{E}^5_\rho \widehat{H}^5_\varphi \right\rangle
	\right) =
		\left\langle \widehat{f^\prime} \widehat{v^\prime} \right\rangle +
		\frac{1}{4}\left\langle \hat f \hat v {\hat w}^2 \right\rangle	.
\label{4_20}
\end{equation}
Here, for brevity, we use the same designations as those given in \eqref{3_a_10} and \eqref{3_b_10}:
\begin{eqnarray}
	\hat{A}^2_t &=& \frac{\hat h}{g}, \quad
	\hat h = \left\langle \hat h \right\rangle + \widehat{\delta h}, \quad
	\left\langle \widehat{\delta h} \right\rangle = 0,
\label{4_30}\\
	\hat{A}^5_z &=& \frac{\hat v}{g}, \quad
	\hat v = \left\langle \hat v \right\rangle + \widehat{\delta v}, \quad
	\left\langle \widehat{\delta v} \right\rangle = 0,
\label{4_40}\\
	\hat{A}^7_\varphi &=& \frac{\rho \hat w}{g}, \quad
	\hat w = \left\langle \hat w \right\rangle + \widehat{\delta w}, \quad
	\left\langle \widehat{\delta w} \right\rangle = 0,
\label{4_50}\\
	\hat{A}^5_t &=& \frac{\hat f}{g} , \quad
	\left\langle \hat{f} \right\rangle = 0 .
\label{4_60}
\end{eqnarray}
According to \eqref{4_40}-\eqref{4_60}, the quantum average in \eqref{4_20} can be written in the form
\begin{eqnarray}
	\left\langle \widehat{f^\prime} \widehat{v^\prime} \right\rangle &=&
	\left\langle \widehat{f^\prime} \, \widehat{\delta v^\prime} \right\rangle ,
\label{4_70}\\
	\left\langle \hat f \hat v {\hat w}^2 \right\rangle &\approx&
	\left\langle \hat f \, \widehat{\delta v} \right\rangle \left\langle\hat{w}\right\rangle^2 .
\label{4_80}
\end{eqnarray}
In classical Proca theory, according to Sec.~\ref{momentum_flux} and Eqs.~\eqref{3_b_100}-\eqref{3_b_130},
the functions $f$ and $v$ are interrelated. It is evident that, in quantum theory, this interrelation persists in the sense that the correlation of quantum fluctuations of the operators
 $\hat f$ and $\widehat{\delta v}$ [the Green function
$
	G_{f, \delta v}(x_1, x_2) =
	\left\langle \hat f(x_1) \widehat{\delta v}(x_2) \right\rangle
$] will be nonzero,
$$
	G_{f, \delta v}(\rho, \rho) =
	\left\langle
		\hat f(\rho) \, \widehat{\delta v}(\rho)
	\right\rangle \neq 0 .
$$
Correspondingly, the Green function \eqref{4_70} will then also be nonzero. Using Eq.~\eqref{4_20}, the momentum density
 takes the form
\begin{equation}
	\left\langle  \widehat{P}^z \right\rangle  \approx
			\left\langle \widehat{f^\prime} \,
		\widehat{\delta v^\prime} \right\rangle +
		\frac{1}{4}\left\langle \hat f \, \widehat{\delta v} \right\rangle \left\langle\hat{w}\right\rangle^2,
\label{4_100}
\end{equation}
and it is nonzero. Notice that the energy flux \eqref{4_20} and the momentum density \eqref{4_100} are of a purely quantum nature.

Thus in this section we have shown that, in quantum Proca theory, there exists a tube with the classical flux of the longitudinal electric field, sourced by `quarks' located at $\pm \infty$,
and with the quantum energy flux/momentum density.

\subsection{Tube with classical energy flux/momentum density and a quantum flux of the longitudinal electric field}
\label{ft_momentum_electric_field}
For this case, we need the following potentials:
$$
	\left\langle \hat{A}^2_t \right\rangle = 0 , \quad
	\left\langle \hat{A}^5_z \right\rangle \neq 0 , \quad
	\left\langle \hat{A}^7_\varphi \right\rangle \neq 0 , \quad
	\left\langle \hat{A}^5_t \right\rangle \neq 0 .
$$
Then, to describe the behavior of the quantum average of the potentials
$\hat{A}^5_t, \hat{A}^5_z,$ and $\hat{A}^7_\varphi$ , one can approximately use Eqs.~\eqref{3_b_100}-\eqref{3_b_130}.

In having these potentials, we can obtain the operator of longitudinal electric field needed for the existence of confinement in the form
\begin{equation}
	\hat E^7_z = \frac{1}{2} \hat{A}^2_t \hat{A}^5_z =
	\frac{1}{2 g} \hat h \hat v .
\label{4_b_20}
\end{equation}
Here, analogous to what was done in Sec.~\ref{ft_electric_field_momentum}, we employ the following designations:
\begin{eqnarray}
	\hat{A}^2_t &=& \frac{\hat h}{g}, \quad
	\left\langle \hat{h} \right\rangle = 0,
\label{4_b_30}\\
	\hat{A}^5_t &=& \frac{\hat f}{g} , \quad
	\hat f = \left\langle \hat f \right\rangle + \widehat{\delta f}, \quad
	\left\langle \widehat{\delta f} \right\rangle = 0,
\label{4_b_40}\\
	\hat{A}^5_z &=& \frac{\hat v}{g}, \quad
	\hat v = \left\langle \hat v \right\rangle + \widehat{\delta v}, \quad
	\left\langle \widehat{\delta v} \right\rangle = 0,
\label{4_b_50}\\
	\hat{A}^7_\varphi &=& \frac{\rho \hat w}{g}, \quad
	\hat w = \left\langle \hat w \right\rangle + \widehat{\delta w}, \quad
	\left\langle \widehat{\delta w} \right\rangle = 0 .
\label{4_b_60}
\end{eqnarray}
According to Eqs.~\eqref{4_b_30} and \eqref{4_b_40}, the quantum average in \eqref{4_b_20} can be written in the form
$$
	\left\langle \hat E^7_z \right\rangle = \frac{1}{2 g}
	\left\langle
		\hat h \, \widehat{\delta v}
	\right\rangle .
$$
Arguments similar to those given in Sec.~\ref{ft_electric_field_momentum} concerning the relationship between
the operators of the potentials  $\hat{A}^2_t = \hat h / g$ and $\hat{A}^5_z = \hat v / g$ which follow from Eqs.~\eqref{3_b_100}-\eqref{3_b_130} can be applied to obtain a Green function
describing the quantum correlation between the operators $\hat{A}^2_t$ and $\hat{A}^5_z$,
$$
	G_{h, \delta v}(\rho, \rho) =
	\left\langle
		\hat h(\rho) \, \widehat{\delta v}(\rho)
	\right\rangle \neq 0 .
$$
Thus the expression for the quantum average of the flux of the longitudinal electric field \eqref{4_b_20} takes the form
$$
	\left\langle \hat \Phi_z \right\rangle =
	2 \pi \int \limits_0^\infty \rho
	\left\langle \hat E^7_z \right\rangle d \rho =
	\frac{\pi}{g} \int \limits_0^\infty \rho
	\left\langle \hat h \, \widehat{\delta v} \right\rangle d \rho	
	< \infty ,
$$
and it is nonzero. Note that this flux between `quarks' located at $\pm \infty$ is of a purely quantum nature.

Thus in this section we have shown that
\begin{itemize}
\item In quantum Proca theory, there can exist tubes possessing simultaneously both the energy flux/momentum density and the flux of the longitudinal electric field.
\item There are two types of such tubes:
	\begin{itemize}
	\item Tubes with (almost) classical flux of the longitudinal electric field and quantum energy flux/momentum density.
	\item Tubes with (almost) classical energy flux/momentum density and a quantum flux of the longitudinal electric field.
	\end{itemize}
\end{itemize}
In the next section, we will show that the presence of tubes between `quarks' in a Proca proton results in the fact that the Proca proton spin includes
a contribution from the momentum density of the Proca field containing inside such tubes. Note that by the Proca proton we mean a hadron in which the interaction between `quarks' is caused by a Proca gluon field.

\section{Angular momentum of flux tubes in the Proca proton
}
\label{Proca_angular_momentum}

Proca proton is an imaginary particle in which a SU(3) gauge field is replaced by a  SU(3) Proca field.
Analogously to an ordinary proton, we assume that there is a tube between `quarks' filled with the Proca field,
which we have considered in Secs.~\ref{ft_electric_field_momentum} and \ref{ft_momentum_electric_field}.

Consider a flux tube connecting two `quarks' (or `quark' and `antiquark') in the Proca proton. We will model such a finite-size tube by cutting it from one of infinite tubes obtained in
Sec.~\ref{ft_electric_field_momentum} or in Sec.~\ref{ft_momentum_electric_field}.

In each tube, there is the following linear momentum density
(for clarity, in this section we resurrect $c$ and $\hbar$ in the equations):
$$
	\mathcal P^i = \frac{2\pi}{ c} \int \limits_0^\infty
	\rho \frac{\epsilon^{i j k}}{\sqrt \gamma} E^a_j H^a_k d \rho,
$$
and this momentum is directed along the axis of each tube.

\begin{figure}[h]
\centering
	\includegraphics[width=0.3\linewidth]{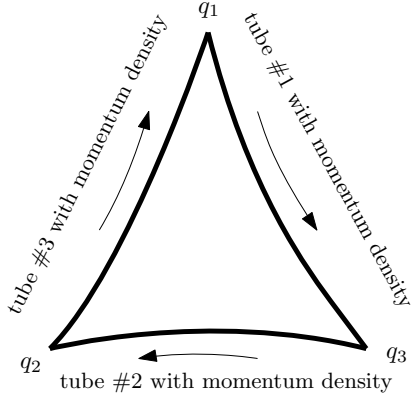}
\caption{The schematic sketch of the Proca proton and tubes with the momentum densities whose directions are shown by the arrows.
}
\label{Proca_proton}
\end{figure}

The schematic sketch of the `quarks', tubes, and directions of momenta is shown in Fig.~\ref{Proca_proton}. It is evident that in this case there is an angular momentum
caused by the presence of the flux of the non-Abelian Proca field. Let us estimate this angular momentum as follows. Each tube has a momentum
$
	\Pi = \mathcal P^z l_\text{tb}
$, where $\mathcal P^z$ is the linear momentum density directed along the tube of the length $l_\text{tb}$, and it is calculated either using the formula
\eqref{3_b_50} for the tube with classical energy flux/momentum density and a quantum  flux of the longitudinal electric field or using the formula
 \eqref{4_100} for the tube with a classical flux of the longitudinal electric field and quantum energy flux/momentum density.

Let us consider the first case. Then the total angular momentum of the non-Abelian Proca field contained in the tubes can be estimated as
$$
	L_g = 3 \Pi r_0 = 3 \mathcal P^z l_\text{tb} r_0 =
	6\pi \frac{l_\text{tb} r_0 \phi^2(0)}{c {g^\prime}^2}
	\int \limits_0^\infty x \left(
		\tilde f^\prime \tilde v^\prime +
		\frac{1}{4}\tilde f \tilde v \tilde w^2
	\right) d x =
	6\pi \frac{l_\text{tb} r_0 \phi^2(0)}{c {g^\prime}^2}
	\tilde \Pi =
	12\pi \frac{\tilde l_\text{tb} \tilde r_0}{{g^\prime}^2} \frac{\hbar}{2} \tilde \Pi ,
$$
where $r_0$ is an estimated value of the distance from the tube to the center of the Proca proton and
$\tilde l_\text{tb} = l_\text{tb} \phi(0)/\sqrt{\hbar c}, \tilde r_0 = r_0 \phi(0)/\sqrt{\hbar c}$, and
${g^\prime}^2 = \hbar c g^2$ are dimensionless quantities.

Thus the fraction of the $\hbar / 2$ Proca proton spin created by the non-Abelian Proca field contained in the tubes carrying the momentum is
$$
	\frac{L_g}{\hbar / 2}  = 12\pi \frac{\tilde l_\text{tb} \tilde r_0}{{g^\prime}^2}
	\tilde \Pi \approx 7.2\,\pi \frac{\tilde l_\text{tb} \tilde r_0}{{g^\prime}^2}.
$$
Here, we have employed the value $\tilde \Pi \approx 0.6$ which has been computed using the solution obtained in Sec.~\ref{momentum_flux} and shown in Fig.~\ref{Poynting}.

One can estimate the magnitude of $L_g / \left(\hbar / 2\right)$ using natural assumptions about the values of the quantities
$\tilde l_\text{tb}$ and $\tilde r_0$. Suppose that, in analogy with quantum chromodynamics, the non-Abelian-Proca-Higgs theory contains
the dimensional parameter $\Lambda_{\text{nAPH}} \approx 1~\text{fm}^{-1}$.
It is therefore natural to assume that
$\phi(0) \approx \sqrt{\hbar c}\,\Lambda_{\text{nAPH}}$. The length of the tube and the distance from the tube to the center can be estimated as
$l_\text{tb} \approx r_0 \approx \Lambda_{\text{nAPH}}^{-1}$. As in quantum chromodynamics, let us also introduce
the dimensionless coupling constant $g^\prime > 1$; then, if one takes, say, $g^\prime = 8.65$, the fraction of the angular momentum created by the Proca gluon field in the Proca proton spin is
$$
	\frac{L_g}{\hbar / 2} \approx 30\% .
$$

Thus in this section we have shown that the Proca theory permits one to explain how the Proca gluon field contributes to the Proca proton spin.

\section{Summary}
\label{concl}

In the present work we have shown that there exist different types of tubes within non-Abelian Proca theory: (a)~the tubes with the flux of the color electric field between `quarks' located at  $\pm \infty$; (b)~the tubes with nonzero momentum density directed along the tube; (c)~the tubes with the classical flux of the electric field and the quantum momentum density; (d)~the tubes with the classical momentum density and the quantum flux of the electric field.

We have considered an imaginary particle~-- Proca proton~-- in which `quarks' are connected by the tubes obtained here.
Using the fact that such tubes may possess a momentum density, we have shown that in this case the angular momentum created by the tubes contributes to the Proca proton spin.

The results can be summarized as follows:
\begin{itemize}
\item We have obtained cylindrically symmetric solutions describing classical tubes possessing either the flux of the longitudinal electric field or the energy flux and the momentum density.
\item At the qualitative level, we have shown that there can exist tubes filled simultaneously both with the flux of the longitudinal electric field and with the energy flux and the momentum density.
But in this case either the flux of the longitudinal electric field or the energy flux/momentum density must be of a purely quantum nature.
 \item We have estimated the contribution of the Proca gluon field, contained in the tubes and creating there the momentum density, to the Proca proton spin.
 \end{itemize}

\section*{Acknowledgments}
We gratefully acknowledge support provided by Grant No.~BR05236730
in Fundamental Research in Natural Sciences by the Ministry of Education and Science of the Republic of Kazakhstan.
We are also grateful to the Research Group Linkage Programme of the Alexander von Humboldt Foundation for the support of this research.

\end{document}